\PassOptionsToPackage{unicode}{hyperref}
\PassOptionsToPackage{hyphens}{url}
\documentclass[
]{article}
\usepackage{amsmath,amssymb}
\usepackage{lmodern}
\usepackage{iftex}
\ifPDFTeX
  \usepackage[T1]{fontenc}
  \usepackage[utf8]{inputenc}
  \usepackage{textcomp} % provide euro and other symbols
\else % if luatex or xetex
  \usepackage{unicode-math}
  \defaultfontfeatures{Scale=MatchLowercase}
  \defaultfontfeatures[\rmfamily]{Ligatures=TeX,Scale=1}
\fi
\IfFileExists{upquote.sty}{\usepackage{upquote}}{}
\IfFileExists{microtype.sty}{% use microtype if available
  \usepackage[]{microtype}
  \UseMicrotypeSet[protrusion]{basicmath} % disable protrusion for tt fonts
}{}
\makeatletter
\@ifundefined{KOMAClassName}{% if non-KOMA class
  \IfFileExists{parskip.sty}{%
    \usepackage{parskip}
  }{% else
    \setlength{\parindent}{0pt}
    \setlength{\parskip}{6pt plus 2pt minus 1pt}}
}{% if KOMA class
  \KOMAoptions{parskip=half}}
\makeatother
\usepackage{xcolor}
\ifLuaTeX
  \usepackage{selnolig}  % disable illegal ligatures
\fi
\IfFileExists{bookmark.sty}{\usepackage{bookmark}}{\usepackage{hyperref}}
\IfFileExists{xurl.sty}{\usepackage{xurl}}{} % add URL line breaks if available
\hypersetup{
  hidelinks,
  pdfcreator={LaTeX via pandoc}}

\author{}
\date{}

\begin{document}

\textbf{Protecting Persona Biometric Data: The Case of Facial Privacy}

\emph{Lambert Hogenhout, Rinzin Wangmo}

\emph{\textbf{Institute for Calculated Futures}}

\textbf{Abstract}

\emph{The proliferation of digital technologies has led to unprecedented
data collection, with facial data emerging as a particularly sensitive
commodity. Companies are increasingly leveraging advanced facial
recognition technologies, often without the explicit consent or
awareness of individuals, to build sophisticated surveillance
capabilities. This practice, fueled by weak and fragmented laws in many
jurisdictions, has created a regulatory vacuum that allows for the
commercialization of personal identity and poses significant threats to
individual privacy and autonomy.}

\emph{This article introduces the concept of Facial Privacy. It analyzes
the profound challenges posed by unregulated facial recognition by
conducting a comprehensive review of existing legal frameworks. It
examines and compares regulations such as the GDPR,
Brazil\textquotesingle s LGPD, Canada\textquotesingle s PIPEDA, and
privacy acts in China, Singapore, South Korea, and Japan, alongside
sector-specific laws in the United States like the Illinois Biometric
Information Privacy Act (BIPA). The analysis highlights the societal
impacts of this technology, including the potential for discriminatory
bias and the long-lasting harm that can result from the theft of
immutable biometric data.}

\emph{Ultimately, the paper argues that existing legal loopholes and
ambiguities leave individuals vulnerable. It proposes a new policy
framework that shifts the paradigm from data as property to a model of
inalienable rights, ensuring that fundamental human rights are upheld
against unchecked technological expansion.}

\textbf{I. Introduction}

The proliferation of digital technologies has ushered in an era of
unprecedented data collection, with facial biometric data emerging as a
particularly sensitive and valuable commodity. Companies are
increasingly leveraging advanced technologies to collect facial data,
often through surreptitious means such as online crawling of publicly
available photos, and subsequently commercializing sophisticated facial
recognition capabilities. This practice, often conducted without the
explicit consent or even awareness of individuals, poses a significant
and escalating threat to personal privacy and autonomy. While initially
marketed to law enforcement and government agencies, the potential for
these capabilities to be sold to commercial entities---such as
department stores, restaurants, and entertainment venues---for real-time
hyper-personalized marketing and pervasive surveillance is rapidly
becoming a reality, transforming personal identity into a tradable asset
within a system that has been defined as ``surveillance capitalism''.
{[}17{]}

The growing threat of unwanted facial data use is multifaceted. Firstly,
\textbf{widespread data collection} by commercial entities, which then
sell facial images and associated personal information, effectively
commodifies an individual\textquotesingle s unique identity without
their consent. Secondly, \textbf{weak privacy laws} in many
jurisdictions, particularly in the United States, have created a
regulatory vacuum that allows these practices to flourish with apparent
impunity. This legal inadequacy means there is little to prevent these
extensive databases from being sold to commercial entities, enabling
targeted advertising and pervasive surveillance. Thirdly, the continuous
advancement of \textbf{AI-powered recognition} technology exacerbates
these concerns, making facial recognition systems more powerful and
accurate each month. This technological progress means that businesses
could soon identify individuals instantly, allowing for the manipulation
of their experiences without their knowledge or explicit permission.

Some jurisdictions are beginning to acknowledge the unique sensitivity
of facial data and the need for \textbf{facial privacy} rights. For
instance, Denmark has recently adopted a law that explicitly recognizes
individuals\textquotesingle{} copyright ownership over their own face
(or facial data). This innovative approach provides a potential model
for safeguarding personal biometric information. However, in many other
countries, including the United States, such explicit protections are
conspicuously absent, leaving individuals vulnerable to the unchecked
exploitation of their facial identity.

This article aims to critically analyze the current landscape of facial
data collection and its profound implications for individual rights and
societal norms. It will provide a brief review of previous academic and
legal discourse surrounding this issue, followed by a comprehensive
comparative analysis of how existing data privacy laws---such as the
General Data Protection Regulation (GDPR), Lei Geral de Proteção de
Dados (LGPD), Personal Information Protection and Electronic Documents
Act (PIPEDA), and relevant laws in China, Singapore, South Korea, and
Japan---address or fail to adequately protect biometric data. Finally,
this article will look forward, identifying the new policies and
recommending a comprehensive policy framework for facial data
protection, or proposing necessary additions and clarifications to
existing legal instruments, to ensure that fundamental human rights are
upheld in the face of rapid technological advancement.

\textbf{II. The Problem: Unconsented Commercialization and Surveillance}

The pervasive nature of facial recognition technology (FRT) and the
increasing commercial value of biometric data have given rise to
significant concerns regarding unconsented commercialization and
surveillance. The core of this problem lies in the methods of data
collection and the subsequent exploitation of this highly sensitive
information.

~~~~\textbf{A. Mechanisms of Data Collection}

Facial data is primarily collected through two main avenues, often
without the explicit knowledge or consent of the individuals concerned:

~~~~~~~~1. \textbf{Web Scraping and Online Crawling:} A prevalent method
involves the automated harvesting of publicly available images from
various online sources, including social media platforms, public
databases, and news articles. Companies like Clearview AI have
notoriously built vast databases of billions of facial images by
scraping the internet, subsequently selling access to these databases to
law enforcement agencies {[}6{]}. This practice blurs the lines between
public and private information, as images posted for social interaction
or news purposes are repurposed for identification and surveillance
without the original intent or consent of the data subject {[}6{]}. The
paper "Data collection via web scraping: privacy and facial recognition
after Clearview" highlights how web scrapers, or bots, systematically
browse and index web pages to download large volumes of data, including
facial images, which are then used to create and operate FRTs {[}6{]}.
The article further emphasizes that public availability of data does not
exempt it from privacy and data protection laws, and that explicit
consent is generally required for such collection and processing
{[}6{]}.

~~~~~~~~2. \textbf{Covert Collection:} Beyond online sources, facial
data is also collected through physical means, often covertly, in public
and semi-public spaces. This includes the deployment of cameras in
retail stores, entertainment venues, and even public streets, which
capture facial images of individuals as they go about their daily lives.
The collected images are then processed to extract unique facial
features, creating biometric templates that can be used for
identification or verification {[}1{]}. This form of collection often
occurs without prominent notice or explicit consent, leading to a sense
of constant surveillance and a significant erosion of individual privacy
{[}1{]}.

~~~~\textbf{B. Commercialization of Facial Data}

The collected facial data, particularly when aggregated into large
databases, becomes a valuable asset for commercialization, leading to
two primary forms of exploitation:

~~~~~~~~1. \textbf{Sale to Law Enforcement and Government Agencies:} The
most immediate and widely reported commercialization involves selling
facial recognition capabilities and access to facial databases to law
enforcement and government entities. While proponents argue this
enhances public safety and national security, critics raise significant
ethical concerns about mass surveillance, potential for misuse, and the
lack of democratic oversight {[}1{]}. The FTC has issued warnings about
the misuses of biometric information, and cases like Rite Aid being
banned from using AI facial recognition technology underscore the
regulatory pushback against unchecked deployment, even in contexts
initially presented as security-enhancing {[}2, 3{]}.

~~~~~~~~2. \textbf{Potential for further Commercial Exploitation:} The
user\textquotesingle s prompt highlights a critical emerging threat: the
sale of facial recognition capabilities to commercial entities beyond
law enforcement. This could include department stores, restaurants, and
entertainment venues. The aim would be \textbf{real-time
hyper-personalized marketing}, where individuals are identified upon
entry, and their past purchasing habits, preferences, or even emotional
states are used to tailor their experience instantly {[}1{]}. This level
of personalized interaction, while potentially convenient for some,
represents a profound intrusion into personal space and autonomy,
turning every public interaction into a targeted marketing opportunity
without consent {[}1{]}---a core tenet of surveillance capitalism
{[}17{]}. The article "Privacy vs. Security: The Legal Implications of
Using Facial Recognition Technology at Entertainment Venues" explicitly
states that FRT is ripe for misuse by businesses, and in the U.S., there
is inadequate protection against companies collecting and using such
data {[}4{]}.

~~~~\textbf{C. Societal and Individual Impacts}

The unconsented collection and commercialization of facial data have
far-reaching implications for individuals and society:

~~~~~~~~1. \textbf{Erosion of Privacy and Autonomy:} The constant threat
of being identified and tracked erodes the fundamental right to privacy
and personal autonomy. Individuals lose control over their own image and
how it is used, leading to a chilling effect on free expression and
association {[}1{]}. The ability to move freely in public without being
subjected to identification and analysis is a cornerstone of democratic
societies, which is undermined by pervasive FRT {[}1{]}.

~~~~~~~~2. \textbf{Risk of Discrimination and Bias:} FRT systems are not
infallible and have been shown to exhibit biases, particularly against
people of color, women, and nonbinary individuals {[}5{]}. These
inaccuracies can lead to false positives or negatives, resulting in
wrongful accusations, misidentification, and discriminatory treatment,
especially when used in critical contexts like law enforcement or access
control {[}1, 6{]}. The ANPD in Brazil and the Privacy Commissioner of
Canada have both highlighted concerns about biases in FRT training
databases and the potential for discrimination {[}7, 8{]}.

~~~~~~~~3. \textbf{Security Risks:} Unlike passwords or credit card
numbers, biometric data like facial features are unique and cannot be
changed if compromised. A data breach involving facial biometric
templates can have irreversible and long-lasting consequences, leading
to identity theft, fraud, and other forms of harm {[}1{]}. The LGPD in
Brazil explicitly classifies biometric data as sensitive, heightening
the risk associated with its processing and emphasizing the need for
robust security measures {[}7{]}. The Irish DPA fined Meta for failing
to adequately protect user data from scrapers, highlighting the risks of
fraud and impersonation {[}6{]}.

These interconnected issues underscore the urgent need for robust legal
and policy frameworks to protect facial biometric data from unconsented
commercialization and surveillance, ensuring that technological
advancements do not come at the cost of fundamental human rights.

\textbf{III. Literature Review: Previous Discourse on Facial Data
Privacy}

The academic and legal communities have increasingly engaged with the
complex issues surrounding facial recognition technology (FRT) and its
implications for privacy. This section reviews key contributions to the
discourse, highlighting the evolution of concerns, central debates, and
the impact of real-world cases.

~~~~\textbf{A. Early Warnings and Academic Concerns}

Early scholarly work on biometric data privacy often focused on the
general risks associated with unique identifiers. However, as FRT
advanced, specific concerns about facial data began to emerge. Articles
like "Beyond surveillance: privacy, ethics, and regulations in face
recognition technology" {[}1{]} have highlighted FRT as a "double-edged
sword," offering societal benefits while simultaneously presenting
complex ethical, legal, and personal challenges. These foundational
pieces emphasized the rapid adoption of FRT in public governance and
security, and the resulting tensions between state/corporate interests
and individual rights {[}1{]}. The discourse evolved from general
surveillance concerns to a more granular examination of commercial
exploitation, particularly as the technology became integrated into
daily life and institutional governance {[}1{]}.

~~~~\textbf{B. Legal and Ethical Debates}

Central to the literature are ongoing legal and ethical debates,
particularly concerning consent, data ownership, and the public nature
of facial images. A significant point of contention is whether data
scraped from publicly available sources can be used without explicit
consent. The article "Data collection via web scraping: privacy and
facial recognition after Clearview" {[}6{]} directly addresses this,
arguing that public availability does not negate privacy rights and that
explicit consent is generally required for the collection and processing
of such data, especially biometric information. This piece also
critiques the inadequacy of existing laws to regulate web scraping
specifically, though it notes that general data protection principles
like those in GDPR can be applied {[}6{]}.

Scholarly critiques frequently point to existing legal loopholes and
enforcement challenges. Many argue that current legislative measures
often fall short of robust scholarly standards and international human
rights norms, necessitating a thorough academic examination beyond
superficial benefits {[}1{]}. The debate also extends to the inherent
biases within FRT systems, with studies showing differential accuracy
rates for various demographic groups, leading to potential
discrimination {[}5{]}. This raises ethical questions about fairness and
the potential for FRT to perpetuate societal inequalities.

~~~~\textbf{C. Case Studies and Real-World Implications}

Real-world cases have significantly shaped the academic discourse,
providing concrete examples of FRT\textquotesingle s implications and
prompting regulatory responses. The "Clearview AI saga" is a prominent
example, extensively analyzed in the literature {[}6{]}. Clearview
AI\textquotesingle s practice of scraping billions of facial images from
the internet to build a database for sale to law enforcement sparked
widespread legal repercussions across multiple jurisdictions. The
article by Lala {[}6{]} details how this case led to enforcement actions
by data protection authorities in Italy, the UK, Australia, and Canada,
highlighting violations of consent requirements, data minimization
principles, and the sensitive nature of biometric data. These cases
underscored the critical need for explicit consent, robust data
protection by design and default, and accountability from data
controllers {[}6{]}.

These real-world applications and their legal challenges have not only
informed public perception but also spurred legislative efforts, such as
the proposed EU AI Act, which seeks to prohibit untargeted scraping of
facial images for FRT databases {[}6{]}. The literature consistently
emphasizes that the irreplaceable nature of biometric data makes data
breaches particularly severe, leading to calls for stringent security
measures and clear accountability frameworks {[}1, 6{]}. The ongoing
academic discussion thus provides a critical foundation for
understanding the multifaceted challenges posed by FRT and for
developing effective policy solutions.

\textbf{IV. Comparative Legal Analysis: Existing Data Privacy Laws and
Biometric Data}

The increasing deployment of facial recognition technology (FRT) has
prompted various jurisdictions worldwide to address the protection of
biometric data within their legal frameworks. This section provides a
comparative analysis of key data privacy laws, examining how they
classify, regulate, and enforce protections for facial biometric data.

\textbf{A. European Union}

The General Data Protection Regulation (GDPR), effective since May 2018,
stands as a cornerstone of data privacy law, significantly influencing
global standards. Under GDPR, facial data is explicitly categorized as
sensitive biometric data under Article 9 {[}9{]}. This classification
imposes stringent requirements for its processing, making it generally
prohibited unless specific conditions are met. These conditions include
the explicit consent of the individual, necessity for employment or
social protection law compliance, protection of vital interests, or
substantial public interest reasons as specified by EU or member state
law {[}9{]}.

Key principles of GDPR, such as data minimization, purpose limitation,
and data protection by design and by default, are particularly relevant
to FRT. Organizations are mandated to collect only necessary data,
process it for specified legitimate purposes, and implement robust
technical and organizational measures to ensure privacy from the outset
{[}9{]}. The GDPR also emphasizes transparency, requiring individuals to
be fully informed about how their data will be used. Enforcement
actions, such as the Information Commissioner's Office (ICO) fines
against Clearview AI, underscore the strict interpretation of GDPR
regarding biometric data, particularly concerning the necessity of
explicit consent and the prohibition of untargeted scraping {[}6, 9{]}.
Implied consent is generally considered insufficient for biometric data
under GDPR {[}9{]}.

\textbf{B. Brazil}

Brazil\textquotesingle s Lei Geral de Proteção de Dados (LGPD), which
came into force in September 2020, similarly classifies biometric data,
including facial data, as sensitive personal data {[}7{]}. This
classification mandates heightened protection and stricter rules for its
processing. The Brazilian Data Protection Authority (ANPD) has actively
engaged with the implications of FRT, releasing reports that analyze the
impacts, risks, and challenges for personal data protection under the
LGPD {[}7{]}.

The ANPD\textquotesingle s guidance highlights several critical
concerns, including the potential for biases in FRT training databases
that could lead to discrimination, and the severe consequences of data
breaches involving biometric templates, which can result in identity
theft and financial fraud {[}7{]}. The LGPD, like GDPR, emphasizes the
need for robust security measures and careful consideration of the
necessity and proportionality of FRT deployment. The proposed Brazilian
Artificial Intelligence Act (Bill No. 2.338/2023) further reinforces
these protections by banning real-time biometric identification systems
in public spaces, with limited exceptions for public security purposes
such as searching for crime victims or in cases of imminent threat
{[}7{]}. This reflects a global trend towards restricting mass
surveillance and ensuring that FRT is not used indiscriminately in
public areas.

\textbf{C. The United States of America}

In contrast to the regional approach in the EU or the state-level
approach in Brazil, data privacy, there is little data privacy
regulation at the federal level in the USA, and what exists is a mosaic
of sector-specific, state-level regulation. There are few specific
references to biometric data or facial data, with the exception of
Illinois' Biometric Information Privacy Act (BIPA) of 2008, which is a
powerful (but geographically limited) model.

BIPA requires informed, written consent and prohibits data collectors
from profiting from Biometric Data.

\textbf{D. Canada}

In Canada, the Personal Information Protection and Electronic Documents
Act (PIPEDA) governs the collection, use, and disclosure of personal
information in the private sector. The Office of the Privacy
Commissioner of Canada (OPC) has issued comprehensive guidance on
processing biometrics under PIPEDA, emphasizing a higher bar for
compliance {[}8{]}. The OPC guidance clarifies that biometric data,
including facial geometry, is considered sensitive by default,
necessitating stringent requirements for consent, safeguards, and
reporting {[}8{]}.

A cornerstone of PIPEDA\textquotesingle s approach to biometrics is a
four-part test that must be satisfied before deployment: legitimate
need, effectiveness, minimal intrusiveness, and proportionality. This
test ensures that biometrics are not used for mere convenience if less
privacy-invasive alternatives exist {[}8{]}. Furthermore, express
consent is the default requirement for biometric data processing. This
means consent must be explicit, informed, specific to biometrics (e.g.,
consent for a photo is not consent for a face embedding), integrated
into user flows, and renewable. Organizations must also provide opt-out
options and alternatives, especially when biometrics are not essential
to their product or service {[}8{]}.

PIPEDA also stresses data minimization and retention limits, requiring
organizations to collect only essential data, favor verification over
identification, store templates under user control where possible, and
destroy records promptly after fulfilling their purpose. The guidance
explicitly states that if an individual withdraws consent, all collected
or created biometric information must be deleted, extending this
requirement to third parties with whom data has been shared {[}8{]}.
Robust safeguards, regular testing for biases and errors, and
transparent accountability mechanisms are also critical components of
compliance under PIPEDA {[}8{]}.

\textbf{E. China}

China has significantly strengthened its data privacy landscape with the
implementation of the Personal Information Protection Law (PIPL), which
came into effect in November 2021, and further detailed regulations
specifically targeting facial recognition technology. The Security
Management Measures for the Application of Facial Recognition
Technology, effective June 1, 2025, introduce stringent rules for FRT
deployment and data handling, aiming to balance technological innovation
with privacy protection {[}10{]}.

Under these regulations, the use of FRT is permitted only when
absolutely necessary and for clearly defined purposes. Organizations
must provide clear justification for its use, adopt the least intrusive
methods, and offer alternatives for identity verification, such as ID
cards {[}10{]}. Notably, the use of FRT is prohibited in sensitive
locations like hotels and public restrooms, and it is illegal to make
FRT mandatory for accessing goods or services if other identification
methods are available {[}10{]}.

PIPL and its accompanying measures impose mandatory transparency
requirements, obliging businesses to provide comprehensive information
to individuals before collecting biometric data. This includes details
about the data processor, purpose of collection, storage duration,
potential impact on rights, and procedures for exercising rights like
withdrawing consent or data deletion. Any changes in data usage must
also be communicated {[}10{]}.

Restrictions on data storage and transfers are also strict: facial
recognition data should ideally remain within the device or system where
it is collected and is generally prohibited from internet transfer
unless legally permitted or with explicit user consent. Retention
periods are limited to the shortest duration necessary. Furthermore,
businesses storing facial recognition records for over 100,000
individuals must register with provincial cyberspace authorities and
submit detailed reports on their data collection, storage, and security
practices {[}10{]}. Enhanced safeguards are also mandated for minors'
facial data, requiring explicit parental or guardian consent and
additional security measures {[}10{]}.

\textbf{F. Other National Laws - Singapore, South Korea and Japan}

\textbf{Singapore}

Singapore's Personal Data Protection Act (PDPA), enforced by the
Personal Data Protection Commission (PDPC), provides a framework for the
collection, use, and disclosure of personal data. The PDPC has issued
specific guidance on the Responsible Use of Biometric Data in Security
Applications, which addresses facial recognition {[}11{]}. Under the
PDPA, biometric data, when associated with other identifying
information, constitutes personal data. Even anonymized biometric
templates require safeguards if there is a possibility of
re-identification {[}11{]}.

The PDPC emphasizes that organizations must obtain consent for the
collection of biometric data unless an exception applies. For security
applications, exceptions can include:

Publicly Available Data: For locations open to the public or where
individuals can be observed by reasonably expected means (e.g., security
cameras), collection without consent is allowed for security purposes
(monitoring or investigations), provided proper notices are displayed
{[}11{]}.

Legitimate Interests: When the legitimate interests of the organization
outweigh any likely adverse effect on the individual, such as installing
security cameras in areas not open to the public or using FRT to monitor
anomalous behavior, a legitimate interests assessment must be conducted,
and reliance on this exception disclosed {[}11{]}. *

Business Improvement: Use of data without consent is permitted if it
allows the organization to improve crowd management or security
operations, provided the purpose cannot be reasonably achieved without
individually identifiable personal data {[}11{]}.

The guidance also addresses risks unique to biometric technology, such
as identity spoofing, errors in identification (false
positives/negatives), and systemic risks to biometric templates. It
recommends measures like anti-spoofing, setting appropriate matching
thresholds, and encrypting biometric templates. Organizations are
advised to minimize personal data storage, process biometric samples
into templates as soon as practicable, and dispose of data permanently
when no longer required {[}11{]}.

\textbf{South Korea}

South Korea's Personal Information Protection Act (PIPA) is a
comprehensive data privacy law that grants individuals significant
rights over their personal information, including the right to be
informed, access, rectify, and erase data {[}12{]}. PIPA classifies
biometric data for the purpose of uniquely identifying someone as a
special class of sensitive information, which necessitates obtaining
separate consent from individuals for its collection and processing
{[}12{]}.

PIPA imposes clear responsibilities on personal information controllers,
including Chief Privacy Officers (CPOs), for protecting and managing
personal information. Online Service Providers face additional
requirements under South Korea's Network Act, including individual
notifications and reporting to the PIPC or Korea Internet \& Security
Agency (KISA) within 24 hours {[}12{]}.

\textbf{Japan}

Japan's primary data protection legislation is the Act on the Protection
of Personal Information (APPI), enforced by the Personal Information
Protection Commission (PPC). While the APPI has been amended to expand
its coverage to include biometric data, it does not currently contain
specific, dedicated rules for biometric data itself {[}13, 14{]}. This
means that while biometric data, including facial images, is recognized
as personal information, its handling falls under the general provisions
of the APPI, rather than a specialized framework {[}13{]}.

However, the PPC has been actively reviewing and proposing stricter
regulations, particularly concerning facial images and the use of facial
recognition systems. Recent proposals indicate a move towards making it
easier for individuals to request the suspension of the use of their
facial-image data {[}15{]}. The PPC's triennial review of the APPI has
also highlighted the need for new rules on biometric data, suggesting
that Japan is moving towards a more explicit and robust regulatory
stance on this sensitive category of information {[}13, 14{]}.

Under the current APPI, operators handling personal information are
generally required to obtain data subjects' consent before providing
personal data to third parties {[}13{]}. However, the absence of
specific rules for biometric data means that the interpretation and
application of consent requirements for facial recognition can be less
clear than in jurisdictions with explicit biometric data
classifications. This gap underscores the ongoing challenge of adapting
general data protection laws to the unique risks posed by advanced
biometric technologies.

\textbf{F. Gaps and Inconsistencies}

The comparative analysis of these diverse legal frameworks reveals
significant progress in addressing biometric data protection, yet also
highlights several critical gaps and inconsistencies that impede
comprehensive safeguarding against the unwanted use of facial data:

Lack of a Unified International Standard: Despite the influence of GDPR,
there is no globally harmonized legal standard specifically for facial
data protection. Similar to the challenges globally operating companies
have see with regard to data privacy in general {[}16{]}, processing of
biometric data suffers from the same patchwork of regulations where the
level of protection varies significantly across jurisdictions. The
absence of a common definition for sensitive biometric data or
consistent consent requirements creates challenges for international
businesses and leaves individuals vulnerable when their data crosses
borders. data privacy in general

Challenges in Cross-Border Enforcement and Jurisdiction: The
internet-driven nature of facial data collection, particularly through
web scraping, often involves entities operating across multiple
jurisdictions. This complicates enforcement efforts, as national data
protection authorities may struggle to assert jurisdiction over foreign
companies or to enforce penalties effectively. The global reach of data
collection often outpaces the territorial limitations of existing laws.

Inadequacy of General Data Protection Laws for Specific Biometric
Threats: While many laws classify biometric data as sensitive, the
specific nuances of facial recognition technology---such as its
potential for mass surveillance, inherent biases, and the irreversible
nature of compromised facial templates---are not always adequately
addressed by general data protection principles. For instance,
Japan\textquotesingle s APPI, while covering biometric data, lacks
specific rules for its handling, which can lead to ambiguities in
application {[}13{]}. Similarly, while consent is a cornerstone, the
practicalities of obtaining truly informed and freely given consent for
pervasive FRT in public spaces remain a significant challenge.

Varying Approaches to Consent and Exceptions: The stringency of consent
requirements differs. GDPR and PIPEDA demand explicit consent, with
limited exceptions, and emphasize alternatives to biometric use {[}8,
9{]}. In contrast, Singapore\textquotesingle s PDPA allows for broader
exceptions based on

legitimate interests or business improvement, which, while requiring
assessments, may offer less stringent protection than explicit consent
{[}11{]}.

Challenges of AI and Machine Learning: The rapid evolution of AI and
machine learning, which underpins advanced FRT, introduces new
complexities. Training AI models often requires vast datasets, and the
origin and consent status of data used for training can be opaque.
Biases embedded in training data can lead to discriminatory outcomes, a
risk acknowledged by the ANPD and OPC {[}7, 8{]}. Existing laws may not
fully address the specific ethical and privacy challenges posed by
AI-driven FRT, such as the inferential capabilities of these systems.

These inconsistencies and gaps underscore the urgent need for a more
cohesive and robust legal framework that specifically addresses the
unique characteristics and risks associated with facial biometric data.

\textbf{V. Policy Recommendations and Future Framework}

The analysis of the current landscape of facial data collection, its
implications, and the existing legal frameworks reveals a pressing need
for a more robust and harmonized approach to protecting persona
biometric data. The following policy recommendations aim to address the
identified gaps and inconsistencies, proposing a comprehensive framework
that prioritizes individual rights and societal well-being. They aim at
shifting power from the data collector to the individual.

\textbf{A. Establishing Facial Data as a Special Category of Personal
Property/Right}

To fundamentally shift the paradigm of facial data ownership and
control, it is imperative to establish facial data as a special category
of personal property or an inalienable personal right. This approach
draws inspiration from innovative legal developments, such as Denmark's
law recognizing copyright ownership over facial data. Granting
individuals explicit control over their facial data would empower them
with fundamental rights, including:

Right to Access: Individuals must have the unequivocal right to access
all facial data collected about them, including biometric templates and
any derived inferences.

Right to Rectification: The ability to correct inaccurate or outdated
facial data, crucial given the potential for bias and misidentification
in FRT systems.

Right to Erasure (Right to be Forgotten): The right to demand the
permanent deletion of their facial data from all databases, especially
when consent is withdrawn or the data is no longer necessary for its
original purpose.

Right to Portability: The ability to obtain and reuse their facial data
for their own purposes across different services, fostering competition
and individual control.

This framework would legally recognize an individual's face as an
extension of their persona, preventing its unauthorized
commercialization and use.

B. Strengthening Consent Mechanisms

Given the sensitive nature of facial biometric data, existing consent
mechanisms often fall short. In particular in the context of large-scale
data scraping where individuals have no opportunity to consent. A
strengthened approach to consent is critical:

Robust, Granular, and Explicit Consent: For all facial data collection
and processing, explicit, informed, and granular consent must be
mandated. This means individuals must be presented with clear,
unambiguous choices regarding specific uses of their facial data, rather
than broad, all-encompassing agreements.

Prohibiting Implied Consent for Biometric Data: Implied consent, where
consent is inferred from an individual's actions, is insufficient for
biometric data. Active, affirmative consent should be required, ensuring
individuals fully understand and agree to the processing of their unique
biological identifiers.

Mandating Clear, Easily Understandable Consent Requests with Opt-Out
Alternatives: Consent requests must be presented in plain language,
avoiding legal jargon, and clearly outline the purpose, scope, and
duration of data processing. Crucially, individuals must always be
offered reasonable and convenient alternatives that do not require the
submission of facial data, especially for accessing essential services.

\textbf{C. Regulating Web Scraping and AI Training Data}

The unchecked practice of web scraping for facial data and its
subsequent use in AI model training represents a significant threat. New
regulations must specifically target these activities:

Prohibiting Untargeted Scraping for Commercial or Surveillance Purposes:
Legislation should explicitly prohibit the mass, untargeted scraping of
facial images from public sources for commercial gain or surveillance
purposes without explicit consent. This would directly address practices
like those employed by Clearview AI.

Requiring Explicit Consent for AI Model Training: Any use of facial data
for training AI models, particularly those intended for FRT, must
require explicit consent from the individuals whose data is used. This
ensures that individuals have a say in how their unique features
contribute to technologies that may later be used to identify or
categorize them.

Implementing Strict Auditing and Transparency Requirements: Developers
and deployers of FRT must be subject to strict auditing and transparency
requirements regarding the datasets used for training. This includes
disclosing the sources of data, methods of collection, and measures
taken to ensure consent and mitigate bias. Independent audits should
verify compliance and assess the fairness and accuracy of trained
models.

\textbf{D. Limiting Commercial Exploitation}

To curb the burgeoning market for facial data, clear limitations on its
commercial exploitation are necessary:

Banning the Sale and Commercial Exchange of Facial Biometric Data: The
outright sale or commercial exchange of raw facial biometric data or
derived templates without explicit, informed, and specific consent
should be prohibited. This would prevent the commodification of personal
identity.

Restricting FRT for Hyper-Personalized Marketing and Profiling in Public
Spaces: The use of FRT for real-time hyper-personalized marketing,
behavioral profiling, or other non-essential commercial purposes in
public or semi-public spaces should be severely restricted or banned.
This protects individuals from constant surveillance and manipulation of
their experiences without their knowledge or consent.

\textbf{E. Enhancing Accountability and Enforcement}

Effective protection requires robust accountability mechanisms and
strong enforcement:

Establishing Independent Regulatory Bodies with Strong Enforcement
Powers: National data protection authorities or newly established
biometric data commissions must be granted sufficient resources,
independence, and powers to investigate, audit, and enforce regulations
effectively. This includes the authority to impose significant penalties
for non-compliance.

Implementing Severe Penalties for Non-Compliance and Data Breaches:
Penalties for violations, particularly those involving sensitive
biometric data, must be substantial enough to act as a genuine
deterrent, reflecting the irreversible nature of biometric data
compromise.

Facilitating Private Rights of Action: Individuals should have clear and
accessible legal avenues to seek redress for violations of their facial
data rights, including the right to compensation for damages incurred
due to unauthorized collection, use, or breach.

\textbf{F. International Cooperation and Harmonization}

The global nature of data flows necessitates international
collaboration:

Developing International Standards and Agreements: Nations must work
towards developing common international standards and agreements for
facial data protection, fostering interoperability between legal
frameworks and ensuring consistent protection across borders.

Promoting Cross-Border Enforcement Mechanisms: International cooperation
among data protection authorities is crucial for effective cross-border
enforcement, particularly against entities engaged in global data
scraping or unauthorized data transfers.

By implementing these recommendations, a comprehensive and
forward-looking policy framework can be established to protect persona
biometric data, ensuring that technological advancements serve humanity
without undermining fundamental rights to privacy and autonomy.

\textbf{VI. Conclusion}

The rapid advancement and pervasive deployment of facial recognition
technology (FRT) have introduced unprecedented challenges to individual
privacy and autonomy, transforming facial biometric data into a
commodity often collected and utilized without consent. This article has
illuminated the escalating threat posed by widespread, unconsented data
collection, the inadequacies of existing privacy laws, and the amplified
risks presented by AI-powered recognition systems. The potential for
commercial exploitation, extending beyond law enforcement to
hyper-personalized marketing and surveillance in everyday commercial
spaces, underscores the urgency of addressing this issue.

Our review of previous academic discourse highlighted a growing concern
within scholarly communities, evolving from general surveillance
anxieties to specific critiques of commercial exploitation and the
ethical dilemmas surrounding data ownership and consent. Prominent
cases, such as Clearview AI, have served as stark reminders of the
real-world implications and the critical need for robust regulatory
responses.

The comparative legal analysis revealed a varied landscape of data
protection. While jurisdictions like the European Union, Brazil, Canada
and China have made strides in classifying biometric data as sensitive
and imposing stricter processing requirements, significant gaps and
inconsistencies persist. These include the absence of a unified
international standard, challenges in cross-border enforcement, and the
inherent limitations of general data protection laws in fully addressing
the unique threats posed by FRT, particularly concerning web scraping
and AI training data. The differing approaches to consent and the
varying degrees of exceptions further complicate the protective
landscape.

In response to these challenges, this article proposes a comprehensive
policy framework designed to safeguard facial biometric data. Key
recommendations include establishing facial data as a special category
of personal property or an inalienable right, thereby granting
individuals explicit control over their unique identity. Strengthening
consent mechanisms through robust, granular, and explicit requirements,
while prohibiting implied consent for biometric data, is paramount.
Furthermore, strict regulation of web scraping and AI training data,
coupled with limitations on commercial exploitation, is essential to
prevent the commodification of personal identity. Finally, enhancing
accountability and enforcement through independent regulatory bodies,
severe penalties, and facilitated private rights of action, alongside
international cooperation, will ensure that these protections are
meaningful and effective.

In an increasingly digitized world, the protection of facial biometric
data is not merely a technical or legal challenge but a fundamental
human rights issue. Balancing technological innovation with individual
privacy and societal well-being requires proactive, comprehensive, and
globally coordinated policy interventions. By adopting the proposed
framework, societies can strive to harness the benefits of FRT while
upholding the dignity, autonomy, and privacy of every individual,
ensuring that our faces remain our own.

\textbf{References}

\begin{enumerate}
\def\labelenumi{\arabic{enumi}.}
\item
  \begin{quote}
  Wang, X., Wu, Y. C., Zhou, M., \& Fu, H. (2024). Beyond surveillance:
  privacy, ethics, and regulations in face recognition technology.
  \emph{Frontiers in Big Data}, \emph{7}, 1337465.
  \url{https://doi.org/10.3389/fdata.2024.1337465}
  \end{quote}
\item
  \begin{quote}
  ACLU Minnesota. (2024, February 29). \emph{Biased technology: The
  automated discrimination of facial recognition}.
  \url{https://www.aclu-mn.org/en/news/biased-technology-automated-discrimination-facial-recognition}
  \end{quote}
\item
  \begin{quote}
  BreachRx. (2022, January 18). \emph{Korea's PIPA becomes one of the
  strictest global privacy laws}.
  \url{https://www.breachrx.com/2022/01/18/south-koreas-personal-information-protection-act/}
  \end{quote}
\item
  \begin{quote}
  China Briefing. (2025, March 31). \emph{China's facial recognition
  regulations: Key business takeaways}.
  \url{https://www.china-briefing.com/news/china-facial-recognition-regulations-2025/}
  \end{quote}
\item
  \begin{quote}
  DLA Piper. (2025, September 2). \emph{Navigating biometrics under
  PIPEDA: OPC\textquotesingle s new guidance raises the bar}.
  \url{https://www.dlapiper.com/en-ca/insights/publications/2025/09/navigating-biometrics-under-pipeda}
  \end{quote}
\item
  \begin{quote}
  Eptura. (2024, June 11). \emph{Before GDPR: Japan's Act on the
  Protection of Personal Information}.
  \url{https://eptura.com/discover-more/blog/before-gdpr-japans-act-on-the-protection-of-personal-information/}
  \end{quote}
\item
  \begin{quote}
  Federal Trade Commission. (2023, May 18). \emph{FTC warns about
  misuses of biometric information and harm to consumers} {[}Press
  release{]}.
  \url{https://www.ftc.gov/news-events/news/press-releases/2023/05/ftc-warns-about-misuses-biometric-information-harm-consumers}
  \end{quote}
\item
  \begin{quote}
  Federal Trade Commission. (2023, December 19). \emph{Rite Aid banned
  from using AI facial recognition after FTC says retailer deployed
  technology without reasonable safeguards} {[}Press release{]}.
  \url{https://www.ftc.gov/news-events/news/press-releases/2023/12/rite-aid-banned-using-ai-facial-recognition-after-ftc-says-retailer-deployed-technology-without}
  \end{quote}
\item
  \begin{quote}
  GDPR Advisor. (n.d.). \emph{GDPR and facial recognition: Privacy
  implications and legal considerations}. Retrieved September 22, 2025,
  from
  \url{https://www.gdpr-advisor.com/gdpr-and-facial-recognition-privacy-implications-and-legal-considerations/}
  \end{quote}
\item
  \begin{quote}
  Hogenhout, L., \& Wang, A. (2023). \emph{Data privacy without
  borders}. Technics Publications.
  \end{quote}
\item
  \begin{quote}
  International Association of Privacy Professionals. (2024, July 11).
  \emph{Japan's DPA publishes interim summary of amendments to data
  protection regulations}.
  \url{https://iapp.org/news/a/japan-s-dpa-publishes-interim-summary-of-amendments-to-data-protection-regulations}
  \end{quote}
\item
  \begin{quote}
  Lala, F. (2023). Data collection via web scraping: privacy and facial
  recognition after Clearview. \emph{i-lex}, \emph{16}(2).
  \url{https://i-lex.unibo.it/article/download/18872/17438}
  \end{quote}
\item
  \begin{quote}
  Mayer Brown. (2024, July 8). \emph{Biometrics and facial recognition -
  ANPD releases report}.
  \url{https://www.mayerbrown.com/en/insights/publications/2024/07/biometrics-and-facial-recognition-anpd-releases-report}
  \end{quote}
\item
  \begin{quote}
  MLex. (2025, February 20). \emph{Japan's PPC proposes stricter data
  protection rules for cookies, facial images, data brokers}.
  \url{https://www.mlex.com/mlex/articles/2300317/japan-s-ppc-proposes-stricter-data-protection-rules-for-cookies-facial-images-data-brokers}
  \end{quote}
\item
  \begin{quote}
  New York State Bar Association. (2025, June 10). \emph{Privacy vs.
  security: The legal implications of using facial recognition
  technology at entertainment venues}.
  \url{https://nysba.org/privacy-vs-security-the-legal-implications-of-using-facial-recognition-technology-at-entertainment-venues/}
  \end{quote}
\item
  \begin{quote}
  Personal Data Protection Commission Singapore. (2022, May 17).
  \emph{Guide on the responsible use of biometric data in security
  applications}.
  \url{https://www.pdpc.gov.sg/help-and-resources/2022/05/guide-on-the-responsible-use-of-biometric-data-in-security-applications}
  \end{quote}
\item
  \begin{quote}
  Zuboff, S. (2019). The age of surveillance capitalism: The fight for a
  human future at the new frontier of power. PublicAffairs.
  \end{quote}
\end{enumerate}

\end{document}